\documentclass[lettersize,journal]{IEEEtran}
\usepackage{amsmath,amsfonts}
\usepackage{algorithmic}
\usepackage{algorithm}
\usepackage{array}
\usepackage[caption=false,font=normalsize,labelfont=sf,textfont=sf]{subfig}
\usepackage{textcomp}
\usepackage{stfloats}
\usepackage{url}
\usepackage{verbatim}
\usepackage{graphicx}
\usepackage{cite}
\newcommand{\vcell}[1]{\begin{tabular}[c]{@{}c@{}}#1\end{tabular}}
\usepackage{cite}

\usepackage{flushend} 

\ifCLASSINFOpdf
\else
  
\fi
\usepackage{multirow}
\usepackage{etoolbox}
\makeatletter
\patchcmd{\@makecaption}
  {\scshape}
  {}
  {}
  {}
\makeatother
\usepackage{xcolor}
\usepackage{hyperref}

\usepackage{array,multirow,graphicx}
\usepackage{float}
\usepackage{comment}
\usepackage{longtable}
\usepackage{tipa}
\usepackage{graphicx}
\usepackage{amsmath}
\usepackage{amssymb}

\usepackage{pifont}
\newcommand{\cmark}{\ding{51}}
\newcommand{\xmark}{\ding{55}}

\begin{document}

\title{A Holistic Link Budget Analysis for mmWave and THz Communications in Non-Terrestrial Networks}

\author{Evla Safahan Ahrazoglu, 
        Eylem Erdogan,
        Ibrahim Altunbas,
        and~Halim Yanikomeroglu\vspace{-0.75cm}
\thanks{Evla Safahan Ahrazoglu and Ibrahim Altunbas are with the Department of Electronics and Communication Engineering, Istanbul Technical University, 34469 Istanbul, Turkey (e-mail: ahrazoglu16@itu.edu.tr; ibraltunbas@itu.edu.tr).}
\thanks{Eylem Erdogan is with the Department of Electrical and Electronics Engineering, Izmir Institute of Technology, 35433 Izmir, Turkey (e-mail: eylemerdogan@iyte.edu.tr).}
\thanks{{Halim Yanikomeroglu is with the Department of Systems and Computer Engineering, Carleton University, ON K1S 5B6 Ottawa, Canada (e-mail: halim@sce.carleton.ca).}}}

\markboth{Journal of \LaTeX\ Class Files,~Vol.~14, No.~8, August~2021}%
{Shell \MakeLowercase{\textit{et al.}}: A Sample Article Using IEEEtran.cls for IEEE Journals}


\maketitle

\begin{abstract}
The non-terrestrial network (NTN) architecture has gained significant interest from the academia owing to its versatility and the ability to provide worldwide service. To achieve extremely high data rates in NTNs, as intended in the sixth-generation (6G) communication systems, millimeter wave (mmWave) and terahertz (THz) frequencies can be considered, enabling substantial bandwidth and data transmission capacity, which makes them highly suitable for NTN applications. However, these high-frequency signals suffer from significant propagation challenges, including atmospheric attenuation, pointing errors, and various environmental effects. Therefore, a comprehensive link budget analysis is essential to accurately assess the feasibility of mmWave/THz-based NTN systems. Existing studies in the literature often fail to fully capture certain frequency-, altitude-, and direction-dependent effects observed in mmWave/THz transmission or possible communication scenarios within the NTN architecture. In particular, while most prior works primarily focus on free-space loss or atmospheric attenuation, this study adopts a much more comprehensive approach. In this work, a detailed link budget analysis is conducted for mmWave/THz NTNs, considering free-space loss, atmospheric absorption, weather-induced effects, ionospheric disturbances, polarization mismatches, feeder losses, antenna and circuitry constraints, fading, pointing errors, and non-white noise characteristics. Here, the variations in the achievable data rate due to operating frequency, altitude, and uplink/downlink transmission are thoroughly examined for all possible communication scenarios including ground-satellite, ground-aerial, aerial-satellite, inter-aerial, and inter-satellite links. The results have revealed that the multi-layer structure of the NTN architecture can help reducing the excessive loss levels to a certain level that can be tolerated by high-gain directional antennas/arrays, providing multi-gigabit links and making mmWave/THz NTNs feasible for 6G communication systems.
\end{abstract}


\section{Introduction}

\IEEEPARstart{W}{ith} the rapid advancements of technology and the growing number of users and data-driven applications, e.g. extended reality \cite{10183792}, smart cities \cite{al2023emerging}, Internet of Things (IoT) \cite{9509294}, etc., the next-generation communication systems aim to offer ultra-reliable links with extreme data rates around the globe \cite{9397776}. To accomplish this, the integration of satellite communication systems and terrestrial networks with the aerial platforms, such as high-altitude platform station (HAPS) systems and uncrewed aerial vehicles (UAVs), can be considered as a promising solution \cite{9915455}.
This multi-layer structure is referred to as non-terrestrial networks (NTNs), and it contributes to the versatility of the sixth-generation (6G) networks while ensuring global connectivity \cite{9380673}.

The NTN architecture comprises of space, aerial, and terrestrial layers. In the space layer, satellites are orbiting at different altitudes, such as low earth orbit (LEO), medium earth orbit (MEO), geostationary orbit (GEO), etc., facilitating a wide range of applications, including television broadcasting, internet, telephony, remote sensing, positioning, {Earth} monitoring, broadband internet, etc. HAPS systems and UAVs are employed in the aerial layer to provide enhanced coverage and reliability in {Earth-space} communications \cite{8438489}. Ground stations and network infrastructures in the terrestrial layer establish the connection with the existing systems, enabling seamless operation. The flexibility introduced by the multi-layer structure of the NTN architecture offers improved disaster recovery, maritime and aeronautical communications as well as real-time applications which require ultra-low latency, {particularly enabled by the low propagation delay of both aerial and low earth orbit satellite links} \cite{9193893}. 
{With the capability of providing service for disastrous, congested, or underserved areas with the NTN architecture, and serving as an alternative to fiber-optic cables for global connectivity, the global coverage with 6G systems becomes attainable \cite{jamshed2024tutorial}.}
{However, to meet the extremely high data rate demands of the next-generation networks, significant enhancements are required in the system capacity. To address this challenge, researchers have explored the use of higher-frequency bands, particularly millimeter wave (mmWave) and terahertz (THz) bands which offer remarkable potential for ultra-high speed communication \cite{9861699,9681870}.}

In mmWave and THz communications, {30-300 GHz and 0.1-10 THz} frequency bands are utilized, respectively, to offer incredibly wide bandwidths and ubiquitous high data rates \cite{8626085,5764977}. Owing to short wavelengths, very high frequencies allow for the use of smaller antennas and improved beamforming, which are additional benefits of these technologies \cite{8663550}. However, mmWave/THz transmission suffers from excessive loss levels mainly caused by the free-space loss and absorption loss. The former arises from the propagation of the radiated electromagnetic wave in a free space, and the latter stems from the interactions between the electromagnetic wave and the absorber particles in the transmission medium \cite{5995306}. In a typical communication scenario, the transmission medium is the atmosphere which contains water vapour and oxygen molecules as the principal absorbers \cite{8568124}. Furthermore, mmWave and THz communications are prone to weather effects, primarily due to rain and cloud/fog formations, as they introduce increased density of the absorber particles in the atmosphere \cite{8826596,electronics12071684}. In addition to these channel characteristics, ions and charged particles also affect the propagation of the electromagnetic wave due to their oscillation, collisions between them, and the Faraday rotation \cite{9541155}. Moreover, additional loss factors, such as the polarization mismatches between transmitter/receiver antennas and loss caused by non-ideal feeder equipment, result in a reduction in the received power, which must be taken into account while designing mmWave and THz systems. To compensate for the extreme loss levels, the utilization of very high-gain antennas/arrays can be considered \cite{10579941}. These transmitter/receiver configurations produce pencil-sharp beams, which require perfect alignment between communicating antennas/arrays. In the presence of misalignment due to non-stable positioning of the communicating nodes, moving platforms, or jitters, the received power significantly diminishes, which is referred to as pointing errors \cite{9931325}. In addition, small-scale fading is also observed in mmWave/THz communications, which originates from the multipath components arriving from non-line-of-sight (NLOS) directions due to scattering caused by the aerosols in the transmission medium \cite{papasotiriou2021experimentally}. {Besides these large- and small-scale effects, {in vertical and slant} mmWave/THz communications, the noise characteristics differ significantly from those in low-frequency communication systems. This difference primarily stems from the frequency-selective nature of atmospheric absorption. Specifically, the absorption affects the microwave brightness temperatures of both the atmosphere  and the Earth, making the noise profile non-white (noise whose power is not evenly distributed across all frequencies) \cite{series2019attenuation}. This frequency selectivity introduces additional challenges in managing the non-white noise in high-frequency communication systems.}
All of the above-mentioned channel effects must be addressed carefully while investigating the feasibility of mmWave/THz {communications in} NTNs since most of them strongly depend on the characteristics of the transmission medium which vary at different altitudes in the atmosphere.

The first altitude-dependent channel effect is the atmospheric absorption loss. Since the densities of the oxygen molecules and water vapour varies with the altitude, the impact of the absorption changes at different layers of the atmosphere \cite{series2019attenuation}. At higher altitudes, the densities of the absorber particles are lower, therefore, the absorption occurs less frequently, resulting in lower loss values. However, the net absorption loss depends both on the altitudes of transmitter and receiver as well as the zenith angle of the link between them. 
Secondly, the impact of weather conditions also depends on the altitude \cite{ITU_rain,ITU_cloudfog}. Since rain and cloud/fog formations are observed in the lowest layers of the atmosphere, ground-ground, ground-aerial, and ground-satellite links are prone to attenuation introduced by adverse weather conditions. Fortunately, inter-aerial, aerial-satellite, and inter-satellite links are not affected by the impacts of rain and cloud/fog formations. 
Moreover, the ionospheric effects can be listed as the third altitude-dependent channel characteristics. Since the densities of ions and charged particles depend on both altitude and the geographical location (latitude and longitude) of the communicating nodes, it is important to take the distribution of ions and charged particles into account while investigating the feasibility of mmWave/THz NTNs \cite{9541155}. For instance, in polar caps (high latitudes), higher ionospheric loss values are observed compared to that in mid-latitude regions \cite{ITU_618_12}. Favorably, the ionospheric loss exhibits an inverse quadratic relationship with the operating frequency, making the mmWave/THz NTNs more robust against the ionospheric effects \cite[Table I]{ITU_618_12}. Lastly, the noise characteristics observed in mmWave/THz NTNs are also altitude-dependent. The noise temperature comprises of the thermal noise of the receiver and the brightness temperature attenuated by the atmosphere \cite{series2019attenuation}. In mmWave/THz NTNs, the brightness temperature can be modeled by the upwelling or downwelling noise temperature in uplink or downlink communications, respectively, making the noise characteristics both altitude- and direction-dependent.

In the literature, several studies that examine the link performance of different mmWave/THz communication scenarios including short range indoor/outdoor, long range terrestrial, and satellite communication systems.
In \cite{6155633}, the achievable data rate is analyzed in the presence of the absorption loss and fog/rain-induced losses for link distances up to 1 km. 
The maximum data rate for a short range THz communication system is examined by accounting for the multipath fading effect in \cite{9569257} for link distances below 60 m.
The maximum data rate and achievable error performance of fixed mmWave and THz communications under extreme weather conditions are studied in \cite{9632553}. 
The link budget evaluation for an objective data rate is performed in \cite{9269930} by taking the modulation order and channel coding rate into account for link distances from 10 m to 1 km.
A few works focus on the link analyses for satellite communication applications. 
In \cite{zhen2018link}, the link budget analysis is performed for an {Earth-satellite} communication system in Tanggula, Tibet, where massive antenna arrays are employed to combat excessive loss levels due to free-space loss, absorption, rain, and cloud. 
The achievable error rate for a THz-enabled intra-orbit link of 1000 km distance in a satellite constellation is evaluated in \cite{10165287}. 
The maximum throughput of a {near-Earth} inter-satellite THz communication link is obtained by accounting for the ionospheric effects, absorption, noise temperature caused by the cosmic microwave background radiation and the sun in \cite{9541155}.
However, the impacts of the aforementioned altitude- and direction-dependent channel effects on the achievable data rate in the NTN architecture {due to weather-induced effects, the atmospheric absorption on the microwave brightness temperatures, and non-white noise} characteristics remain unexplored in the literature. 

Therefore, in this study, a thorough investigation on the feasibility of mmWave/THz NTNs is presented by taking all of the aforementioned varying channel characteristics into consideration. The contributions of this paper are summarized as follows:
\begin{itemize}
\item Different from the existing studies on link budget analysis regarding THz systems mentioned above, this work simultaneously analyzes all possible communication scenarios within the multi-layered structure of the NTN architecture, including both uplink and downlink communications for ground-satellite, ground-aerial, aerial-satellite, inter-aerial, and inter-satellite links scenarios.
\item The altitude-dependent atmospheric absorption, weather-induced attenuation (rain, cloud, and fog), and non-white noise characteristics driven by upwelling and downwelling brightness temperatures are incorporated together with the most recent advances in the mmWave/THz electronics.
\item Practical channel constraints, such as ionospheric effects, polarization mismatches, feeder losses, multipath fading, and pointing errors, are taken into account.
\item {The impact of antenna and array configurations is analyzed to provide practical guidelines for high-gain design, specifically examining how different antenna types (lens, horn, and cuboid) and physical aperture sizes influence robustness against pointing errors. Also, sensitivity analysis is included to examine the robustness of different antenna and array configurations against the orientation fluctuation level.}
\item The achievable data rate is obtained for all possible communication scenarios within the NTN architecture to quantify the performance. The results have revealed that the multi-layer structure of the NTN architecture provides significant advantages in utilizing the mmWave and THz bands to meet 6G requirements. 
\item {The outage probability analysis is performed. It is shown via both theoretical and simulations that in mmWave/THz NTNs, the statistics of the pointing error becomes the determining factor.} 
\item Comprehensive analyses and simulations demonstrate that mmWave and THz links can be established regardless of atmospheric conditions in inter-aerial, inter-satellite, and aerial-satellite links, owing to the thinner atmosphere and high-gain antennas/arrays. In contrast, both weather and atmospheric conditions reduce the achievable data rate in ground-based links. 
\end{itemize}
The detailed comparison of this work with closely related studies in the literature can be found in Table \ref{table_lit}.






\begin{table*}[!t]
\centering
\caption{Comparison of This Work with the Related Link Budget Analyses in the Literature}
\label{table_lit}
\resizebox{\linewidth}{!}{%
\begin{tabular}{|l|c|c|c|c|c|c|c|c|}
\hline
 & \cite{9541155} 
 & \cite{6155633} 
 & \cite{9569257} 
 & \cite{9632553} 
 & \cite{9269930} 
 & \cite{zhen2018link} 
 & \cite{10165287} 
 & This work \\ 
\hline

Absorption 
& not required${}^*$ 
& \cmark 
& \cmark 
& \cmark 
& \cmark 
& \cmark 
& not required${}^*$ 
& \cmark \\ 
\hline

Weather effects 
& not required${}^*$ 
& \cmark 
& \xmark 
& \cmark 
& \xmark 
& \cmark 
& not required${}^*$ 
& \cmark \\ 
\hline

Ionospheric effects 
& \cmark 
& \xmark 
& \xmark 
& \xmark 
& \xmark 
& \xmark 
& \xmark 
& \cmark \\ 
\hline

Additional losses 
& \xmark 
& \xmark 
& \cmark 
& \xmark 
& \xmark 
& \xmark 
& \cmark 
& \cmark \\ 
\hline

Small-scale effects${}^{**}$ 
& PE 
& \xmark 
& F 
& F \& PE 
& F \& PE 
& \xmark 
& \xmark 
& F \& PE \\ 
\hline

\shortstack{Frequency-dependent\\noise characteristics}
& \xmark 
& \xmark 
& \xmark 
& \xmark 
& \xmark 
& \cmark 
& \xmark 
& \cmark \\ 
\hline

Frequency bands 
& mmWave/THz 
& mmWave/THz 
& THz 
& mmWave/THz 
& THz 
& THz 
& THz 
& mmWave/THz \\ 
\hline

Considered scenarios 
& \vcell{inter-satellite}
& \vcell{terrestrial}
& \vcell{indoor}
& \vcell{terrestrial}
& \vcell{short/long\\range terrestrial}
& \vcell{ground-satellite}
& \vcell{inter-satellite}
& \vcell{uplink/downlink\\ground-satellite\\ground-aerial\\aerial-satellite\\inter-aerial\\inter-satellite} \\ 
\hline

\multicolumn{9}{l}{${}^{*\textcolor{white}{*}}$ Since inter-satellite links are considered in these studies, the absorption loss and weather effects are not included.} \\

\multicolumn{9}{l}{${}^{**}$ In this row, F and PE indicate whether the fading and pointing error effects are considered, respectively.}

\end{tabular}%
}
\end{table*}

The rest of the paper is organized as follows: The main channel effects are introduced and elaborated in Section II. The noise characteristics are explained by accounting for the variations due to frequency, altitude, and transmission direction, and the achievable data rate is evaluated in Section III. The numerical results are presented in Section IV for uplink/downlink transmission in ground-aerial, ground-satellite, aerial-satellite, inter-aerial, and inter-satellite links. Finally, Section V concludes the paper.

\section{Main Channel Effects and Transceiver Requirements}

\label{Sec.5.I}

To analyze the feasibility of {mmWave}/{THz} {NTNs}, the main channel effects should be modeled by accounting for the varying characteristics of the transmission medium with respect to altitude, atmospheric conditions, and weather events. In the following, the large-scale channel effects, such as the free-space loss, atmospheric absorption loss, attenuation due to weather conditions, ionospheric effects, polarization mismatch between antennas, and feeder losses, as well as the small-scale channel effects, such as the fading and pointing errors, are presented. Table \ref{table_sym} summarizes all the parameters used in the paper. 

\begin{table}[!t]
\centering
{
\caption{List of Symbols}
\label{table_sym}
\resizebox*{1\linewidth}{!}{
\begin{tabular}{|c|l|}
\hline
Symbol & Definition\\ \hline
$f_{\rm GHz}$ & Frequency in GHz \\ \hline
$\lambda$ & Wavelength  \\ \hline
$W$ & System bandwidth \\ \hline
$d$ & Link distance  \\ \hline
$L^{\rm fsl}$ & Free-space loss  \\ \hline
$L^{\rm abs}$ & Atmospheric absorption loss  \\ \hline
$\varphi^{\rm abs}$ & Specific attenuation due to absorption  \\ \hline
$L^{\rm wea}$ & Loss due to weather conditions  \\ \hline
$L^{\rm rain}$ & Loss due to rain  \\ \hline
$\mathcal{R}$ & Rainfall rate  \\ \hline
$k_{\rm rain}$, $\rho_{\rm rain}$ & Frequency dependent constants for rain attenuation  \\ \hline
$L^{\rm cf}$ & Loss due to cloud/fog formations  \\ \hline
$\varphi^{\rm cf}$ & Specific attenuation due to cloud/fog formations  \\ \hline
$\rho_{\rm cf}$ & Liquid water density  \\ \hline
$\mathcal{K}'_{\rm cf}$ & Liquid water specific attenuation coefficient  \\ \hline
$\mathcal{K}_{\rm cf}$ & Cloud liquid mass absorption coefficient  \\ \hline
$V$ & Visibility range  \\ \hline
$\mathcal{W}_{\rm cf}$ & Integrated liquid water content  \\ \hline
$L^{\rm add}$ & Additional loss due to polarization mismatch and feeder losses  \\ \hline
$S_{n}$ & Noise power spectral density  \\ \hline
$T_{0}$ & Receiver thermal noise temperature  \\ \hline
$T_{\rm Br}$ & Brightness temperature  \\ \hline
$\mathcal{F}_{n}$ & Receiver's noise figure  \\ \hline
$\mathcal{G}^{\rm q}$ & Array gain  \\ \hline
$\mathcal{G}_{\rm e}^{\rm q}$ & Antenna element gain \\ \hline
\end{tabular}}}
\end{table}


\subsection{Free-Space Loss}

Once the electromagnetic wave is transmitted from an antenna, the transmitted power is distributed on a sphere. The power intensity on a unit area on this sphere decreases as the electromagnetic wave propagates. Thus, for a fixed receiver aperture, the received power reduces for larger link distances. This phenomenon is called the free-space loss or spreading, and it has a well-known mathematical model given by the Friis formula as \cite[Sec. 2.17.1]{balanis2016antenna}
\begin{equation}
L^{\rm fsl} = \left(\frac{4 \pi d}{\lambda}\right)^{2},
\label{Eq.5.0.001}
\end{equation}
\noindent where $d$ and $\lambda$ are the link distance and the operating wavelength, respectively. Here, it can be seen that the free-space loss increases quadratically as the operating wavelength gets smaller or the transmission distance gets larger. {It is important to note that the free-space loss is applicable {at any operating frequency} to all types of links in the NTN architecture, including inter-layer (e.g. ground-aerial, ground-satellite, and aerial-satellite) and intra-layer (e.g. aerial-aerial and satellite-satellite) scenarios.}

\subsection{Atmospheric Absorption Loss}
At high frequencies, the electromagnetic wave interacts with the absorber particles in the transmission medium, leading to molecular absorption loss. In NTNs, the transmission medium is the atmosphere which contains the oxygen molecules and water vapour as the principal absorbers. The specific attenuation due to absorption up to 1000 GHz is given as \cite{series2019attenuation}
\begin{equation}
\varphi^{\rm abs} 
= 0.182 f_{\rm GHz} \left(N''_{\rm oxy}(f_{\rm GHz}) + N''_{\rm wv}(f_{\rm GHz})\right)\hspace{3mm}\text{[dB/km]},
\label{Eq.5.0.002}
\end{equation}
\noindent where $f_{\rm GHz}$ is the operating frequency in GHz. Moreover, $N''_{\rm oxy}(f_{\rm GHz})$ and $N''_{\rm wv}(f_{\rm GHz})$ are the imaginary parts of the frequency-dependent complex refractivities of the oxygen molecules and water vapour in the atmosphere, respectively, {which are related to the atmospheric parameters, such as pressure, humidity level, and temperature which can be found in \cite{series2019atmosphere} or can be extracted from the local data if available.} {$N''_{\rm oxy}(f_{\rm GHz})$ and $N''_{\rm wv}(f_{\rm GHz})$ are evaluated according to the corresponding shape factor and strength of the absorption lines around the frequency of interest as described in \cite{series2019attenuation}.}
Since the densities of the absorber particles are constant along horizontal paths, the atmospheric absorption loss is obtained by
\begin{equation}
L^{\rm abs} = \varphi^{\rm abs}d \hspace{3mm}\text{[dB]}.
\label{Eq.5.0.003}
\end{equation}
\noindent On the other hand, for slant paths, the densities of the absorber particles vary along the propagation path. Thus, the atmospheric absorption loss can be approximately computed by dividing the atmosphere into exponentially increasing layers as
{
\begin{equation}
L^{\rm abs} \approx \sum_{i = i_{\rm lower}}^{i_{\rm upper}} \varphi^{\rm abs}_{i} d_{i} \hspace{3mm}\text{[dB]},
\label{Eq.5.0.004}
\end{equation}}
\noindent where $i_{\rm lower}$ and $i_{\rm upper}$ are the indexes of the lowest and the uppermost layers of the atmosphere in which the transmitter and receiver are located. Furthermore, $\varphi^{\rm abs}_{i}$ and $d_{i}$ are the corresponding specific attenuation [dB/km] and the path length [km] of the $i$-th layer, respectively. Please refer to \cite{series2019attenuation} for the detailed calculation of $\varphi^{\rm abs}_{i}$ and $d_{i}$.

\begin{figure}[!t]
\centering
 	\resizebox*{1\linewidth}{!}{\includegraphics{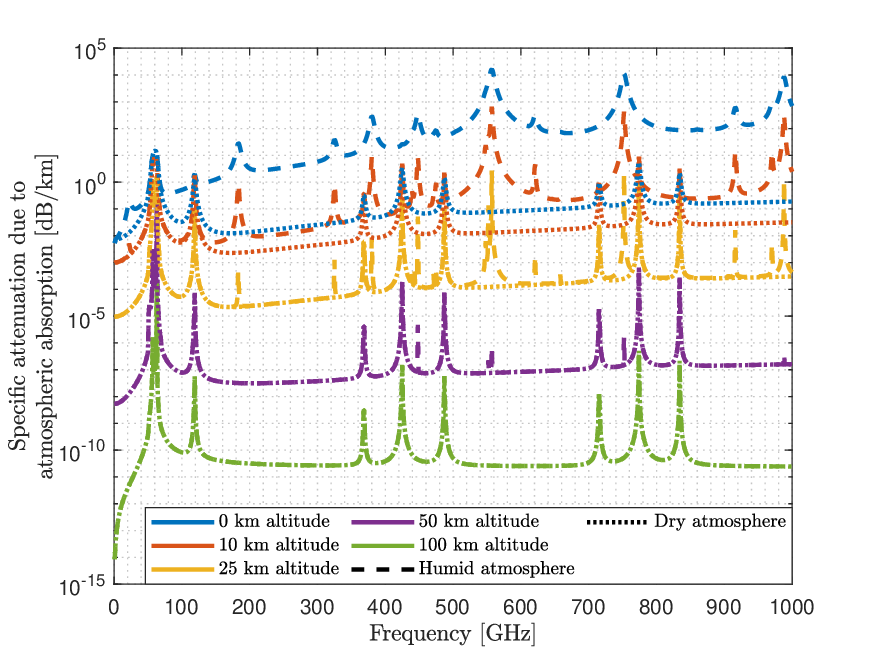}}
  \vspace{-20pt}
\caption{Specific attenuation due to absorption at different altitudes for dry and humid standard atmospheric conditions.}
\label{fig.5.0.001}
\end{figure}

In Fig. \ref{fig.5.0.001}, the specific attenuation at different altitudes are shown for dry and humid {standard atmospheric conditions} \cite{series2019atmosphere}. {As observed, the atmosphere introduces absorption peaks at certain frequencies, permitting signal transmission only within the communication windows between these peak values.}
Moreover, significant difference is observed in the specific attenuation values under dry and humid atmospheric conditions at lower altitudes, which stems from the fact that the oxygen molecules and water vapour lie mostly in the lowest layers of the atmosphere, leading to increased absorption loss in humid air. At higher altitudes, the curves for dry and humid atmospheric conditions get closer to each other. Furthermore, it can be deduced that some absorption peaks disappear at higher altitudes, e.g. at 183 GHz, 325 GHz, 557 GHz, 916 GHz, enabling larger transmission windows. Additionally, it can be inferred from the figure that the absorption loss is negligible at high altitudes, which makes the mmWave/THz transmission feasible for NTNs, especially in aerial-to-satellite, inter-aerial, and inter-satellite links.

\subsection{Loss due to Weather Cconditions}
In addition to the free-space loss and atmospheric absorption loss, the performance of mmWave and THz communications can be degraded by the adverse weather conditions in the lower layers of the atmosphere, {especially in the ground-aerial and ground-satellite links}, such as rain, clouds, and fog formations, because of the increased density of the absorber particles.

In the presence of rain, the specific attenuation due to rain {up to 1000 GHz} is written as $\varphi^{\rm rain} = k_{\rm rain}\mathcal{R}^{\rho_{\rm rain}}$ [dB/km], where $k_{\rm rain} \in \{k_{\rm H},k_{\rm V},k_{\rm C}\}$ and $\rho_{\rm rain} \in \{\rho_{\rm H},\rho_{\rm V},\rho_{\rm C}\}$ are frequency-dependent constants which can be found in \cite{ITU_rain}, and $\mathcal{R}$ is the rainfall rate. Here, $k_{\rm H}$ and $\rho_{\rm H}$ {stand for the constants for horizontal polarization}, whereas $k_{\rm V}$ and $\rho_{\rm V}$ {denote the constants for vertical polarization}. For circular polarization, $k_{\rm C}$ and $\rho_{\rm C}$ are given as \cite{ITU_rain}
\begin{equation}
\begin{split}
k_{\rm C} &= \frac{k_{\rm H} + k_{\rm V} + \left(k_{\rm H}-k_{\rm V}\right)\cos^2\theta\cos2\varpi}{2},\\
\rho_{\rm C} &= \frac{ k_{\rm H}\rho_{\rm H} + k_{\rm V}\rho_{\rm V} + \left(k_{\rm H}\rho_{\rm H} - k_{\rm V}\rho_{\rm V}\right)\cos^2\theta\cos2\varpi }{2k_{\rm C}},
\label{Eq.5.0.005}
\end{split}
\end{equation}
\noindent {where $\theta$ is the elevation angle of the path, and $\varpi$ is the polarization tilt angle ($\varpi = 45^\circ$ for circular polarization).} By using the specific attenuation, the total attenuation due to rain can be found by $L^{\rm rain} = \varphi^{\rm rain} d_{\rm eff}$, where $d_{\rm eff}$ is the effective path length under rainy weather conditions. 

{
In the presence of cloud/fog formations, the specific attenuation can be obtained as $\varphi^{\rm cf} = \mathcal{K}'_{\rm cf} \rho_{\rm cf}$ [dB/km] {up to 1000 GHz} \cite{ITU_cloudfog,11000031}. Here, the liquid water density is given as $\rho_{\rm cf} = (18.35 V)^{-1.43}$ [g/m$^3$] for advection fog and $\rho_{\rm cf} = (42 V)^{-1.54}$ [g/m$^3$] for radiation fog \cite{liao2023attenuation}, where $V$ [km] denotes the visibility. Additionally, cloud liquid water specific attenuation coefficient is written as $\mathcal{K}_{\rm cf}' = \frac{0.819 f_{\rm GHz}}{\varepsilon''(1+\eta^{2})}$ [(dB/km)/(g/m$^3$)], where $\eta = \frac{2 + \varepsilon'}{\varepsilon''}$. In this formulation, $\varepsilon''$ and $\varepsilon'$ are given as 
\begin{equation}
\begin{split}
\varepsilon'' &= \frac{f_{\rm GHz} \left(\varepsilon_{0}-\varepsilon_{1}\right)}{f_{\rm p}\left[1+\left(f_{\rm GHz} / f_{\rm p}\right)^{2}\right]} + \frac{f_{\rm GHz}\left(\varepsilon_{1}-\varepsilon_{2}\right)}{f_{\rm s}\left[1+\left(f_{\rm GHz} / f_{\rm s}\right)^{2}\right]},\\
\varepsilon' &= \frac{\varepsilon_{0}-\varepsilon_{1}}{\left[1+\left(f_{\rm GHz} / f_{\rm p}\right)^{2}\right]}+\frac{\varepsilon_{1}-\varepsilon_{2}}{\left[1+\left(f_{\rm GHz} / f_{\rm s}\right)^{2}\right]}+\varepsilon_{2},
\label{Eq.5.0.007}
\end{split}
\end{equation}
\noindent where $\varepsilon_{0} = 77.66 + 103.3\left(\frac{300}{T}-1\right)$, $\varepsilon_{1} = 0.0671\varepsilon_{0}$, $\varepsilon_{2} = 3.52$, $f_{\rm p} = 20.20 - 146\left(\frac{300}{T}-1\right) + 316\left(\frac{300}{T}-1\right)^{2}$, $f_{\rm s} = 39.8 f_{\rm p}$, and $T$ is the liquid water temperature in Kelvin \cite{ITU_cloudfog}. For slant paths, as considered in ground-aerial and ground satellite links, the attenuation due to cloud/fog can be found as $L^{\rm cf} = \frac{\mathcal{K}_{\rm cf} \mathcal{W}_{\rm cf}}{\sin\theta}$ [dB] \cite{ITU_cloudfog}, where $\mathcal{W}_{\rm cf} = \delta_{\rm cf}\rho_{\rm cf} \times 10^{-3}$ [kg/m$^{2}$] is the integrated liquid water content, $\delta_{\rm cf}$ is the thickness of cloud/fog, and $\theta$ is the elevation angle. Here, the cloud liquid mass absorption coefficient $\mathcal{K}_{\rm cf}$ [dB/(kg/m$^2$)] is defined as \cite{ITU_cloudfog}
\begin{equation}
\footnotesize
\begin{split}
\mathcal{K}_{\rm cf} = \mathcal{K}_{\rm cf}' \left(0.1522 e^{-\frac{(f_{\rm GHz}-f_{1})^{2}}{\sigma_{1}}} + 11.51 e^{-\frac{(f_{\rm GHz}-f_{2})^{2}}{\sigma_{2}}} - 10.4912\right),
\label{Eq.5.0.006}
\end{split}
\end{equation}
for $T = 273.75^\circ$C, where $f_{1} = -23.9589$, $f_{2} = 219.2096$, $\sigma_{1} = 3.2991 \times 10^{3}$, and $\sigma_{2} = 2.7595 \times 10^{6}$. By taking both attenuation due to rain and cloud/fog formations into account, the loss factor due to weather conditions can be written as $L^{\rm wea} = L^{\rm rain}L^{\rm cf}$.
}

\begin{figure}[!t]
\centering
 	\resizebox*{1\linewidth}{!}{\includegraphics{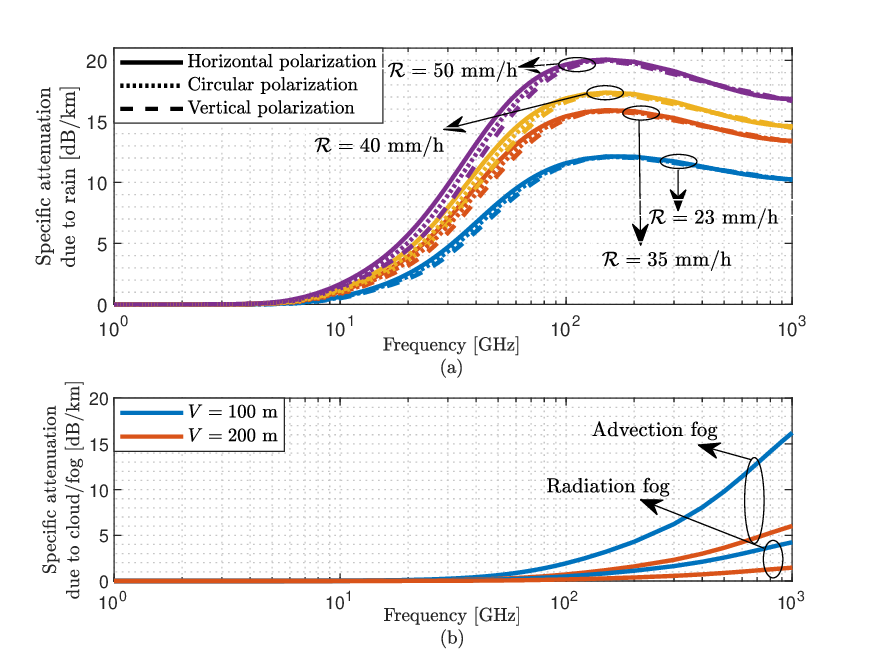}}
  \vspace{-20pt}
\caption{(a) Specific attenuation due to rain for different rainfall rates. (b) Specific attenuation due to cloud/fog for different visibility ranges, $T = 10^\circ$C and $90^\circ$ elevation angle.}
\label{fig.5.0.002}
\end{figure}

In Fig. \ref{fig.5.0.002}a, the specific attenuation due to rain is illustrated from 1 GHz to 1000 GHz for horizontal, vertical, and circular polarizations under several rainfall rates. {At frequencies {below} 10 GHz, the impact of rain on signal attenuation is minimal. However, as the frequency increases, significant attenuation becomes noticeable reaching its peak at 100 GHz.}
In addition, it is inferred from the figure that attenuation is almost the same for different polarizations over the whole frequency range. {In Fig. \ref{fig.5.0.002}b, specific attenuation due to cloud/fog formations is shown for different visibility ranges and fog types. Similar to the rainy weather case, cloud/fog formations have negligible impact for lower frequencies. Above 100 GHz operating frequency, however, cloud/fog formations can significantly reduce the received power. Also, it can be seen that lower visibility ranges lead to {severer} attenuation due to thicker fog. }
Since both rain and cloud/fog formations are observed at lower altitudes up to a few km above ground, it can be concluded that the communication links between ground and aerial nodes, as well as those between ground and satellites, are prone to adverse weather conditions.

\subsection{Ionospheric Effects}

{
The mmWave/THz transmission is also affected by the ions and charged particles in the ionosphere. Two metrics can be used to describe the ionospheric attenuation, namely, the plasma frequency and the collision frequency \cite{9541155}. The electromagnetic waves lead free electrons and ions to oscillate while propagating through plasma. This phenomenon is characterized by the plasma frequency which can be calculated as
\begin{equation}
f_{\rm pla} = \frac{q_{\rm el}}{2 \pi}\sqrt{\frac{N_{\rm el}}{\epsilon_{0} m_{\rm el}}},
\label{Eq_R1_C2_01}
\end{equation}
where $q_{\rm el} = 1.6021\times 10^{-19}$ Coulombs is the electron charge, $m_{\rm el} = 9.109 \times 10^{-31}$ kg denotes the mass of a single electron, $N_{\rm el}$ [m$^{-3}$] represents the electron density in the ionosphere, and $\epsilon_{0} = 8.854 \times 10^{-12}$ F/m shows the vacuum permittivity \cite{1006342}.
At very high frequencies, the impact of plasma frequency diminishes, which can be interpreted as the free electrons and ions falling short compared to the frequency of the electromagnetic wave \cite{9541155}. {In addition to the above-given metric}, the collisions among charged particles, ions, electrons, and neutral particles may cause attenuation. The collision frequency measures the intensity of these interactions and is given as
\begin{equation}
f_{\rm col} = f_{\rm ei} + f_{\rm en} + f_{\rm in}.
\label{Eq_R1_C2_02}
\end{equation}
Here, $f_{\rm ei}$, $f_{\rm en}$, and $f_{\rm in}$ are related to the collisions of electrons with ions, electrons with neutral particles, and ions with neutral particles, respectively, and they are expressed as \cite{9541155}
\begin{equation}
\begin{split}
f_{\rm ei} &= \left(34 + 4.18 \log_{10}\left(\frac{T_{\rm ion}^{3}}{N_{\rm el}'}\right)\right) N_{\rm el}' T_{\rm ion}^{-3/2},\\
f_{\rm en} &= 5.4 \times 10^{-10} N_{\rm ne}' T_{\rm ion}^{1/2},\\
f_{\rm in} &= 2.6 \times 10^{-9} (N_{\rm el}' + N_{\rm ne}) \mathcal{W}_{\rm in}^{-1/2}.\\
\end{split}
\label{Eq_R1_C2_03}
\end{equation}
In this formulation, $T_{\rm ion}$ is the temperature with the average value of 1000 K in the ionosphere. Moreover, $\mathcal{W}_{\rm in} = 28.96$ {is the} molecular weight of ions and charged particles (assumed equal). Also, $N_{\rm el}'$ and $N_{\rm ne}$ are the densities of the electrons and neutral particles in cm$^{-3}$ \cite{1006342}.

The ionospheric attenuation per km can be calculated based on the plasma frequency and the collision frequency as \cite{1006342}
\begin{equation}
{\footnotesize
\begin{split}
\varphi^{\rm ion} = \frac{k}{2} \left(\frac{f_{\rm pla}}{f_{\rm GHz} \times 10^{9}}\right)^{2} \frac{f_{\rm col}}{f_{\rm GHz} \times 10^{9}} \left(1 + \frac{1}{2} \left(\frac{f_{\rm pla}}{f_{\rm GHz} \times 10^{9}}\right)^{2} \right),
\end{split}}
\label{Eq_R1_C2_04}
\end{equation}
where $k$ is the wave number. The ionospheric attenuation can  be calculated as $L^{\rm ion} = \varphi^{\rm ion} d_{\rm ion}$, where $d_{\rm ion}$ is the path length in the ionosphere. Here, the densities of the electron and neutral particles can be extracted from \cite{bilitza2018iri}. 

\begin{figure}[!t]
\centering
 	\resizebox*{1\linewidth}{!}{\includegraphics{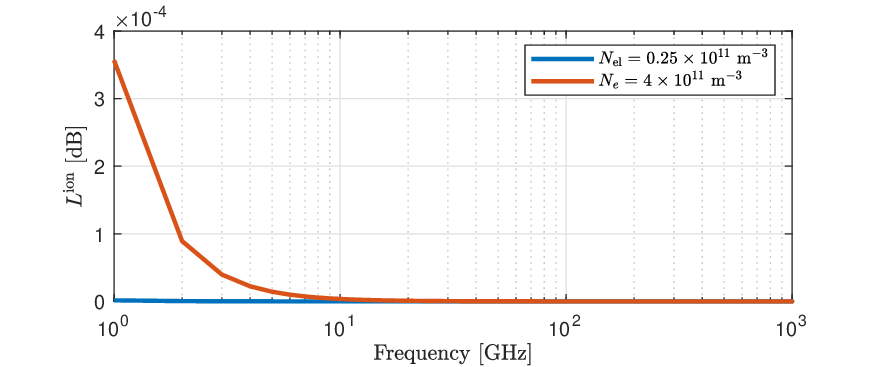}}
    \vspace{-20pt}
\caption{{Ionospheric loss values from 1 GHz to 1000 GHz operating frequencies.}}
\vspace{-4mm}
\label{fig_R1_2_1}
\end{figure}
In Fig. \ref{fig_R1_2_1}, the ionospheric attenuation is illustrated for two different values of the electron density, i.e. $N_{\rm el} = 0.25 \times 10^{11}$ m$^{-3}$ for low electron density and $N_{\rm el} = 4 \times 10^{11}$ m$^{-3}$ for high electron density, and for the density of the neutral particles, a peak value is chosen as $N_{\rm el} = 3 \times 10^{5}$ cm$^{-3}$ \cite{kelley2009earth}. Here, the path length in the ionosphere is assumed as $d_{\rm ion} = 940$ km as the ionosphere layer is approximately in between 60 km and 1000 km altitudes. As can be seen from the figure, the ionospheric attenuation rapidly diminishes with the increasing operating frequency. Above 10 GHz, the ionospheric attenuation is less than $10^{-4}$ dB. Similar results are reported in ITU recommendations as the ionospheric attenuation reduces quadratically {as the frequency increases} \cite[Table III]{ITU_531_15}, and they are insignificant above 10 GHz \cite[Table I]{ITU_618_12}. Since the objective of this work is to analyze the link performance of mmWave/THz NTNs, which operate above 30 GHz, the ionospheric effects are considered as insignificant.}


\subsection{Additional Loss Factors}

Apart from the signal attenuation as described in the previous subsections, communication links in NTNs may also suffer from other losses due to polarization mismatch between antennas, feeder losses etc. The polarization mismatch occurs when the transmitter and receiver antennas do not have the same polarization, which inevitably reduce the received signal power. Moreover, the feeder loss may arise from the cables, connectors, and impedance mismatch between the connected components. In order to take these factors into account, $L^{\rm add}$ {is taken into consideration in the link budget analysis} {with typical values ranging between 1-3 dB \cite{5447021}}.

{
\subsection{Antenna Requirements}

As explained so far, several attenuation factors are present in mmWave/THz communications, reducing the received power and degrading the system performance. To compensate for the excessive loss levels, highly directional antennas/arrays are employed. By utilizing antenna arrays, higher gains at both the transmitter (t) and receiver (r) can be achieved, enabling long range links. 
{For an $N \times N$ uniform planar array of antennas with the maximum gain of $\mathcal{G}_{\rm e}^{\rm q}$, the maximum array gain can be expressed as \cite{10841410}
\begin{equation}
\begin{split}
\mathcal{G}^{\rm q} = N^{2} \mathcal{G}_{\rm e}^{\rm q}, \hspace{3pt}  \text{for} \hspace{3pt} \rm q \in \{t,r\}.
\label{Eq.5.0.008}
\end{split}
\end{equation}
In the case of $\frac{\lambda}{2}$ element spacing, the physical aperture can be written as $\ell \times \ell = \frac{N \lambda}{2} \times \frac{N \lambda}{2}$. For fixed $\ell \times \ell$, the number of antennas can be obtained as $N^{2} = \left(\lfloor \frac{2 \ell}{\lambda} \rfloor\right)^{2}$, where $\lfloor \cdot \rfloor$ is the floor operator. For a fixed array gain and a fixed antenna element gain, the number of antennas can be found as $N^{2} = \lceil \frac{\mathcal{G}^{\rm q}}{\mathcal{G}_{\rm e}^{\rm q}} \rceil$, where $\lceil \cdot\rceil$ is the ceiling operator.} 
It can be inferred here that the maximum array gains at both the transmitter and receiver can be increased {proportionally to the square of the decreasing wavelength} by keeping the physical aperture of the array fixed. Thus, the losses caused by spreading, atmospheric absorption, and adverse weather conditions can be tolerated.

}

{
\subsection{Fading and Pointing Error}
\label{subsec2G}
Besides loss factors, wireless channels experience rapid fluctuations in the received signal's amplitude, referred to as the small-scale effects, which can be decomposed as multipath fading and pointing errors. In mmWave/THz frequencies, fading primarily emerges from scattering due to the aerosols present in the transmission medium. This results in multipath components of significant power being detected by the receiver even if they arrive from {NLOS} directions \cite{10263616}. {In the literature, the $\alpha$-$\mu$ distribution is widely adopted to model the fading channel coefficient, denoted as $h_{\rm f}$, since it is capable of modeling the sparse channel characteristics where the strong LOS connection exhibits minor fading effect. Hence, $h_{\rm f}$ can be characterized by the following probability density function (PDF) \cite{4067122}:}
\begin{equation}
\begin{split}
f_{h_{\rm f}}(x) = \frac{\alpha\mu^\mu x^{\alpha\mu-1}
}{\hat{h}_{\rm f}^{\alpha\mu}\Gamma(\mu)}\exp\left(-\mu{x^\alpha}/{\hat{h}_{\rm f}^{\alpha}}\right), \hspace{3pt} 0 \leq x,
\label{Eq.5.0.008_TVT}
\end{split}
\end{equation}
where $\alpha$ and $\mu$ are the distribution parameters, $\hat{h}_{\rm f} = \sqrt[\alpha]{{E}\left[h_{\rm f}^\alpha\right]}$ is the $\alpha$-root mean value of $h_{\rm f}$, and $E[\cdot]$ is the expectation operator. 

In addition to the multipath fading, the misalignment between the transmitter and receiver can lead to small-scale variations, which originates from the utilization of highly directional antennas/arrays to combat the excessive loss levels in mmWave/THz communications. Since these antennas/arrays employ pencil-sharp beams, they are very susceptible to imperfect alignment caused by the relative movements of the communicating nodes and/or the antenna jitters, leading to a reduction in the received power also known as pointing error \cite{9931325}. {In a recently proposed pointing error model for mmWave/THz communications, the impact of antenna and array designs are incorporated in the statistics of the pointing error coefficient, denoted as $h_{\rm p}$, and the PDF of $h_{\rm p}$ is given as {\cite[Eq. (10a)]{10841410}}
\begin{equation}
\begin{split}
f_{h_{\rm p}}(x) = -\psi^{2}x^{\psi-1}\ln x, \hspace{3pt} 0 \leq x \leq 1,
\label{Eq.5.0.009_TVT}
\end{split}
\end{equation}
where $\psi = {\sigma^{2}}/{\sigma_{\theta}^2}$, and $\sigma_{\theta}$ is the standard deviation of the orientation fluctuation \cite{10841410}. In this formulation, the beam divergence of the array pattern is expressed as $\sigma =  \sqrt{\frac{\sigma_{\rm e}^{2}\sigma_{\rm a}^{2}}{\sigma_{\rm a}^{2} + k^{2}\sigma_{\rm e}^{2}}}$ \cite{10841410}, where $\sigma_{\rm e} \approx 0.6 \theta_{{\rm e},3{\rm dB}}$ denotes the beam divergence of the antenna element \cite[Eq. (8.1.12)]{vaughan2003channels}, $\sigma_{\rm a} = k\sqrt{{10}/{(b \ln 10)}} \sin \left({ \theta_{{\rm a},3{\rm dB}}}/{2}\right)$ is the beam divergence of the array factor \cite{8080225}, $k = 2 \pi / \lambda$ is the wave number, and $b = 3$. Here, $\theta_{{\rm e},3{\rm dB}}$ is the half power beamwidth (HPBW) of the antenna element and can be extracted from recent antenna designs at THz frequencies, whereas $\theta_{{\rm a},3{\rm dB}} = 2\left|\frac{\pi}{2} - \frac{\lambda}{2\pi d_{\rm e}}\frac{2.782}{N}\right|$ is the HPBW of the uniform planar array with element spacing $d_{\rm e}$ {\cite[Eq. (6.14c)]{balanis2016antenna}.}}


}

{
\subsection{Noise Characteristics}
In the most of the existing studies, the noise observed at the receiver is modeled as a thermal noise. However, in mmWave/THz bands, the noise characteristics differ from the thermal noise model because of the cosmic microwave brightness temperature, the microwave brightness temperature of the Earth, and attenuation due to atmospheric absorption. {As a result, uplink and downlink communication links exhibit distinct noise profiles \cite{series2019attenuation}}.
{Specifically, downlink communications (e.g. aerial-to-ground, satellite-to-ground, satellite-to-aerial,  links) are affected by the downwelling brightness temperature, while uplink communications (e.g. ground-to-aerial, ground-to-satellite, aerial-to-satellite links) are influenced by the upwelling brightness temperature}.
The downwelling brightness temperature $T_{\rm Br,down}$ is the sum of cosmic microwave brightness temperature attenuated by the atmospheric absorption and downwelling atmospheric microwave brightness temperature \cite[Sec. 4.1]{series2019attenuation}. {On the contrary,} the upwelling brightness temperature $T_{\rm Br,up}$ is slightly different as it consists of the upwelling  atmospheric microwave brightness temperature, the downwelling atmospheric microwave brightness temperature reflected by the Earth’s surface attenuated by the net atmospheric absorption, and upwelling microwave brightness temperature of the Earth’s surface attenuated by the atmospheric absorption \cite[Sec. 4.2]{series2019attenuation}. Once the brightness temperature is calculated as shown in \cite{series2019attenuation}, the single-sided noise power spectral density can be written as
\begin{equation}
S_{n} = k_{\rm B} (T_{\rm 0} + T_{\rm Br}),
\label{Eq.5.0.0.009}
\end{equation}
\noindent where $T_{0}$ is the receiver thermal noise temperature and $k_{\rm B}$ represents the Boltzmann's constant. Here, $T_{\rm Br} = T_{\rm Br,down}$ for downlink and $T_{\rm Br} = T_{\rm Br,up}$ for uplink communications.
\begin{figure}[!t]
\centering
 	\resizebox*{1\linewidth}{!}{\includegraphics{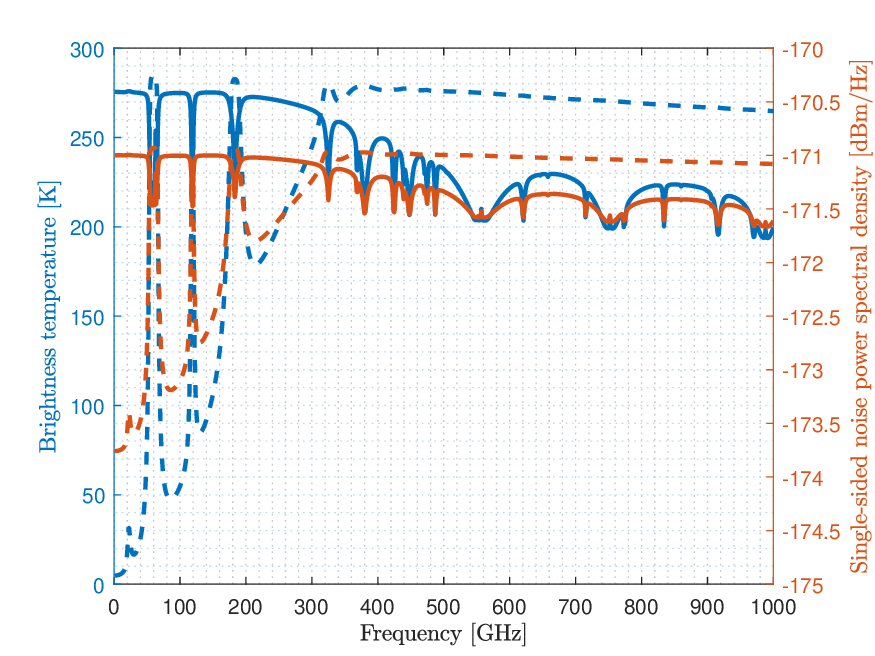}}
    \vspace{-20pt}
\caption{{Upwelling (solid line) and downwelling (dashed line)} brightness temperatures and the single-sided noise power spectral density for a link between 0 km and 100 km altitudes with $0^\circ$ zenith angle.}
\label{fig.5.0.003}
\end{figure}
In Fig. \ref{fig.5.0.003}, the brightness temperature and the single-sided noise power spectral density are illustrated {for $T_{0} = 300$ K} for a link between 0 km and 100 km altitudes with zenith angle of $0^\circ$ under humid atmospheric conditions. It can be deduced from the figure that the brightness temperature varies with operating frequency {due to} the absorption affected on the cosmic microwave brightness temperature and the brightness temperature of the Earth's surface. Thus, the noise power is non-white. {For the analysis, it is assumed that the system bandwidth, denoted as $W$, is divided into small sub-bands of width $\Delta W$ and that the noise power spectral density is constant within each sub-band.}
Hence, the noise power observed in each sub-band can be expressed as {\cite{zhen2018link}}
\begin{equation}
P_{{n},l} = k_{\rm B} (T_{\rm 0} + T_{\rm Br}) \Delta W \mathcal{F}_{n},
\label{Eq.5.0.0.010}
\end{equation}
\noindent where $\mathcal{F}_{n}$ is the receiver's noise figure and $l$ is the sub-band index. Throughout this work, $\Delta W = 100$ MHz is considered. After obtaining the noise power in uplink/downlink communications, achievable data rate can be found for various communication scenarios in mmWave/THz NTNs.

}

By taking the free-space loss, atmospheric absorption loss, weather-dependent losses, polarization mismatch, feeder losses, arrays gains, link margin, and the noise characteristics into consideration, the received power and the achievable data rate can be evaluated for a wide range of communication scenarios, which is elaborated in the next section.

{{

\section{Performance Analyses and Engineering Insights}

\subsection{Received Power and Achievable Data Rate}

Recalling that the system bandwidth is divided into small sub-bands of width $\Delta W$, 
the received power in each sub-band is evaluated as
\begin{equation}
P_{{\rm r},l} = \frac{P_{{\rm t},l} \mathcal{G}^{\rm t} \mathcal{G}^{\rm r} h_{{\rm f},l}^{2} h_{{\rm p},l}^{2}}{L^{\rm fsl}_{l} L^{\rm abs}_{l} L^{\rm wea}_{l} L^{\rm ion}_{l} L^{\rm add}_{l}},
\label{Eq.5.0.0.011}
\end{equation}
\noindent where $P_{{\rm t},l}$ is the transmit power of the $l$-th sub-band, and it is assumed that $P_{{\rm t}}$ is equally divided across each sub-band.} In this formulation, {all loss values are} calculated for the center frequency of the $l$-th sub-band as shown in Section \ref{Sec.5.I}. 
{
Therefore, the achievable data rate can be expressed as the sum of the average achievable data rates in each sub-band as {\cite[Eq. (15)]{zhen2018link}}
{
\begin{subequations}
\begin{align}
C &= \sum_{l} E\left[\Delta W \log_{2}\left(1 + \frac{P_{{\rm r},l} }{P_{{n},l}}\right)\right] \label{Eq.5.0.0.012a}\\
&\leq \sum_{l} \Delta W \log_{2}\left(1 + \frac{P_{{\rm t},l}}{L^{\rm tot}_{l} P_{{n},l}}\right) \label{Eq.5.0.0.012b},
\end{align}
\label{Eq.5.0.0.012}
\end{subequations}}

\noindent {where $L^{\rm tot}_{l} = L^{\rm fsl}_{l} L^{\rm abs}_{l} L^{\rm wea}_{l} L^{\rm ion}_{l} L^{\rm add}_{l}/\left(E[h_{{\rm f},l}^{2}] E[h_{{\rm p},l}^{2}]\right)$ is the total loss which includes all large- and small-scale effects}.} In \eqref{Eq.5.0.0.012b}, the inequality is obtained by employing the Jensen's inequality and the independence of $h_{{\rm f},l}$ and $h_{{\rm p},l}$. Here, the average fading channel gain is found as $E[h_{{\rm f},l}^{2}] = \frac{\hat{h}_{\rm f}^{2}}{\mu ^{2/\alpha}} \frac{\Gamma(\frac{2}{\alpha} + \mu)}{\Gamma (\mu)}$ \cite[Eq. (102)]{8610080}, and the squared expected value of $h_{{\rm p},l}$, i.e. $E[h_{{\rm p},l}^{2}] = \int_{0}^{1} x^{2} f_{h_{\rm p}}(x) dx$, can be found as 
\begin{equation}
\begin{split}
E[h_{{\rm p},l}^{2}] =  -\psi^{2}\int_{0}^{1} x^{\psi+1}\ln x dx.
\end{split}
\label{Eq.5.0.0.012_TVT0}
\end{equation} 
By applying integration by parts in the above-equation, $E[h_{{\rm p},l}^{2}]$ can be rewritten as
\begin{equation}
\begin{split}
E[h_{{\rm p},l}^{2}] &= \psi^{2} \int_{0}^{1}\frac{x^{\psi + 1}}{\psi + 2} dx = \frac{\psi^{2}}{(\psi + 2)^{2}}.
\end{split}
\label{Eq.5.0.0.012_TVT1}
\end{equation} 
{By substituting $E[h_{{\rm f},l}^{2}]$ and $E[h_{{\rm p},l}^{2}]$ {in total loss expression then} in \eqref{Eq.5.0.0.012b}, the upper bound for the achievable data rate can be obtained for all possible inter-layer and intra-layer communication scenarios in mmWave/THz NTNs as
{\begin{equation}
{\scriptsize
\begin{split}
C \leq \sum_{l} \Delta W \log_{2}\left(1 + \frac{P_{{\rm t},l} \mathcal{G}^{\rm t} \mathcal{G}^{\rm r}    }{L^{\rm fsl}_{l} L^{\rm abs}_{l} L^{\rm wea}_{l} L^{\rm ion}_{l} L^{\rm add}_{l} P_{{n},l}}         \frac{\hat{h}_{\rm f}^{2} \Gamma(\frac{2}{\alpha} + \mu) \psi^{2}}{\mu ^{2/\alpha} \Gamma (\mu) (\psi + 2)^{2}} \right).
\end{split}} 
\label{Eq.5.0.0.012_TVT2}
\end{equation}}}

Alternatively, the fading and pointing error can also be taken into consideration by defining a link margin $L^{\rm mar}_{l}$ and replacing the denominator $(h_{{\rm f},l}^{2}h_{{\rm p},l}^{2})$ in the total loss expression by $\frac{1}{L^{\rm mar}_{l}}$ for a simpler analysis.}


{
\subsection{Outage Probability}

In the considered setup, the outage probability can be defined as probability that the {signal-to-noise ratio (SNR)} of at least one sub-band falling below a predefined threshold $\gamma_{\rm th}$. Mathematically, it can be formulated as
\begin{equation}
\begin{split}
P_{\rm out} = \Pr\left[\min_{l}\gamma_{l} \leq \gamma_{\rm th}\right] = 1 - \prod_{l} (1 - F_{\gamma_{l}}(\gamma_{\rm th})),
\end{split}
\label{Eq.5.0.0.012_out_TVT1}
\end{equation} 
where $\gamma_{l} = \frac{P_{{\rm r},l} }{P_{{n},l}}$ is the SNR of the $l$-th sub-band. Here, it is assumed that SNRs in different sub-bands are statistically independent. In the above-equation, the {cumulative distribution function (CDF)} of $\gamma_{l}$ can be written in terms of the CDF of the joint channel coefficient $h_{{\rm fp},l} = h_{{\rm f},l}h_{{\rm p},l}$ by using \eqref{Eq.5.0.0.011} as 
\begin{equation}
\begin{split}
F_{\gamma_{l}}(x) = F_{h_{{\rm fp},l}}\left(\sqrt{\frac{xL^{\rm fsl}_{l} L^{\rm abs}_{l} L^{\rm wea}_{l} L^{\rm ion}_{l} L^{\rm add}_{l}}{P_{{\rm t},l} \mathcal{G}^{\rm t} \mathcal{G}^{\rm r}}}\right),
\end{split}
\label{Eq.5.0.0.013_out_TVT1}
\end{equation} 
where the CDF of the joint channel coefficient $h_{{\rm fp},l}$ can be obtained by using \eqref{Eq.5.0.008_TVT} and \eqref{Eq.5.0.009_TVT} as \cite[Eq. (19)]{10841410}
\begin{equation}
{\small
\begin{split}
F_{h_{{\rm fp},l}}(x) &= \frac{1}{\Gamma(\mu)} \gamma\left(\mu,\frac{\mu}{\hat{h}_{\rm f}^{\alpha}} x^{\alpha}\right)  + \frac{\psi c  \mu^{\frac{\psi}{\alpha}} x^\psi}{\Gamma(\mu) \hat{h}_{\rm f}^{\psi}}  \Gamma\left(\mu-\frac{\psi}{\alpha},\frac{\mu}{\hat{h}_{\rm f}^{\alpha}} x^{\alpha}\right) \\
&- \frac{\psi^2 c}{\psi + \frac{1}{c}}\frac{ \mu^{\frac{\psi}{\alpha}+\frac{1}{\alpha c}} x^{\psi+\frac{1}{c}}}{\Gamma(\mu) \hat{h}_{\rm f}^{\psi+\frac{1}{c}}}  \Gamma\left(\mu-\frac{\psi}{\alpha}-\frac{1}{\alpha c},\frac{\mu}{\hat{h}_{\rm f}^{\alpha}} x^{\alpha}\right).
\end{split}}
\label{Eq025}
\end{equation}
Here, $c$ is a large number. Hence, the outage probability can be derived as
\begin{equation}
{
\begin{split}
P_{\rm out} &= 1 - \prod_{l} \Bigg[1 - \frac{1}{\Gamma(\mu)} \gamma\left(\mu,\frac{\mu}{\hat{h}_{\rm f}^{\alpha}} A^{\alpha}\gamma_{\rm th}^{\alpha/2}\right)\\  
&- \frac{\psi c  \mu^{\frac{\psi}{\alpha}} A^\psi \gamma_{\rm th}^{\psi/2}}{\Gamma(\mu) \hat{h}_{\rm f}^{\psi}}  \Gamma\left(\mu-\frac{\psi}{\alpha},\frac{\mu}{\hat{h}_{\rm f}^{\alpha}} A^{\alpha}\gamma_{\rm th}^{\alpha/2}\right) \\
&+ \frac{\psi^2 c}{\psi + \frac{1}{c}}\frac{ \mu^{\frac{\psi}{\alpha}+\frac{1}{\alpha c}} A^{\psi+\frac{1}{c}} \gamma_{\rm th}^{\frac{\psi}{2}+\frac{1}{2c}}}{\Gamma(\mu) \hat{h}_{\rm f}^{\psi+\frac{1}{c}}} \\ 
& \times  \Gamma\left(\mu-\frac{\psi}{\alpha}-\frac{1}{\alpha c},\frac{\mu}{\hat{h}_{\rm f}^{\alpha}} A^{\alpha}\gamma_{\rm th}^{\alpha/2}\right)\Bigg],
\end{split}}
\label{Eq.5.0.0.014_out_TVT1}
\end{equation} 
where $A = \sqrt{{L^{\rm fsl}_{l} L^{\rm abs}_{l} L^{\rm wea}_{l} L^{\rm ion}_{l} L^{\rm add}_{l}}/{(P_{{\rm t},l} \mathcal{G}^{\rm t} \mathcal{G}^{\rm r})}}$.}

{
\subsection{Engineering Insights}


In this subsection, the feasibility assessment of the mmWave/THz NTNs is performed by considering the achievable data rates and outage probability. The altitudes of the ground, aerial, and satellite nodes are assumed as 0 km, 30 km, and 700 km above mean sea level, respectively. }
Unless otherwise stated, $0^\circ$ zenith angle (or equivalently $90^\circ$ elevation angle) is considered for all inter-layer communication links. Moreover, the transmission bandwidth is taken as the $2\%$ of the operating frequency. Furthermore, relying on the most recent advances in the mmWave and THz electronics, the transmit power of $P_{{\rm t}} = 20$ dBm (\hspace{-0.25pt}\cite{9086884,9675289,9431537} and references therein) and noise figure of $\mathcal{F}_{n} = 10$ dB (\hspace{-0.25pt}\cite{7452667,5337876,7593272} and references therein) are considered. The additional loss $L_{\rm add} = 3$ dB is assumed to account for the feeder losses and polarization mismatch. {In addition, {unless otherwise stated,} standard uniform antenna arrays of size $N \times N$ with element spacing $d_{\rm e} = \lambda/2$ are used at both transmitter and receiver with the antenna elements described in \cite{3GPP_rep} with maximum element gain of $\mathcal{G}_{\rm e}^{\rm q} = 8$ dB and HPBW of $65^\circ$ for $\rm q \in \{t,r\}$.} {Throughout the section, the theoretical curves are obtained by the upper bound in \eqref{Eq.5.0.0.012b}, whereas the simulation curves are computed according to \eqref{Eq.5.0.0.012a} by generating $10^{4}$ realizations of the fading and pointing error coefficients.} {Throughout the simulations, the fading coefficients are generated by the $\alpha$-root of a Gamma distributed random variable {with shape parameter of $\mu$ and scale parameter of $\frac{1}{\mu}$}, whereas the pointing error coefficients are generated by the negative exponent of a Gamma distributed random variable (with shape parameter of 2 and scale parameter $\psi$ as described in Section \ref{subsec2G}).} The parameters that are used in this section are summarized in Table \ref{table_param}.

\begin{table}[!t]
\centering
{
\caption{System Parameters}
\label{table_param}
\resizebox*{1\linewidth}{!}{
\begin{tabular}{|l|l|}
\hline
Parameter & Value\\ \hline
Satellite altitude & 700 km \cite{8700141}  \\ \hline
Aerial altitude & 30 km \cite{radioregulations}  \\ \hline
Ground altitude & 0 km   \\ \hline
System bandwidth $W$ & 2\% of the operating frequency \\ \hline
Transmit power $P_{\rm t}$ & 20 dBm \cite{9086884,9675289,9431537} \\ \hline
Receiver's noise figure & 10 dB \cite{7452667,5337876,7593272} \\ \hline
Additional loss factor $L^{\rm add}$ & 3 dB \\ \hline
Antenna element gain $\mathcal{G}_{\rm e}^{\rm q}$ & 8 dBi \cite{3GPP_rep} \\ \hline
HPBW of the antenna element & $65^\circ$ \cite{3GPP_rep} \\ \hline
Atmospheric conditions & Standard dry and humid \cite{series2019atmosphere} \\ \hline
Rainfall rate & $\mathcal{R} = 0, 23, 35, 40, 50$ mm/h \cite{ITU_rain} \\ \hline
Visibility range in cloud/fog formations & $V = 50$ m \cite{ITU_cloudfog} \\ \hline
\end{tabular}}}
\end{table}



\begin{figure}[!t]
\centering
 	\resizebox*{1\linewidth}{!}{\includegraphics{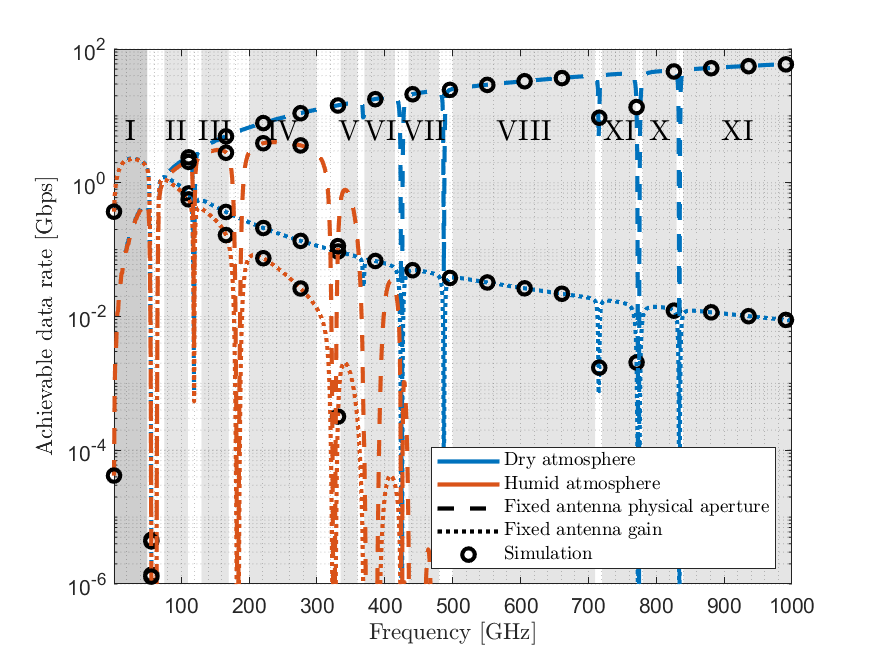}}
    \vspace{-20pt}
\caption{{Achievable data rate in ground-to-satellite link for fixed physical aperture ($25 \times 25$ cm${}^{2}$) and fixed gain (50 dBi) under dry and humid atmospheric conditions.}}
\label{fig.5.0.004}
\end{figure}

In Fig. \ref{fig.5.0.004}, the achievable data rate is illustrated up to 1 THz for a ground-to-satellite link under dry/humid atmospheric conditions. Here, fixed physical aperture ($25 \times 25$ cm${}^{2}$) and fixed gain (50 dBi) arrays are used {with fading parameters $\alpha = 6$ and $\mu = 5$ and standard deviation of the orientation fluctuation $\sigma_{\theta} = 0.001^\circ$}. As can be seen, {the theoretical curves provide very tight upper bound for the simulations.} {Additionally, the figure clearly illustrates the transmission windows between the absorption peaks labeled from I to XI. It is evident that frequency bands from I-IV (which align with transmission windows around 60 GHz, 120 GHz, 180 GHz, and 330 GHz) can provide acceptable achievable data rates in humid atmosphere, whereas in the dry atmospheric conditions, all frequency bands can be exploited under fixed antenna physical aperture assumption.} Moreover, it is evident that fixed array gain outperforms fixed physical aperture below 120 GHz operating frequency. {At high} frequencies, the achievable data rate for fixed array gain tends to reduce, and fixed physical aperture provides enhanced data rates, which originates from fact that the realized gain increases with the increasing frequency for a fixed physical aperture. Furthermore, in dry air, the atmospheric absorption loss levels are significantly lower compared to those in humid air. Thus, fixed physical aperture can tolerate the loss values under dry atmospheric conditions as the operating frequency increases. However, the absorption loss is incredibly high under humid atmospheric conditions due to the increased densities of the absorber particles in the lower layers of the atmosphere. Hence, both fixed physical aperture and fixed array gain perform poorly above 435 GHz operating frequency in ground-to-satellite links in humid atmospheric conditions. 

\begin{figure}[!t]
\centering
 	\resizebox*{1\linewidth}{!}{\includegraphics{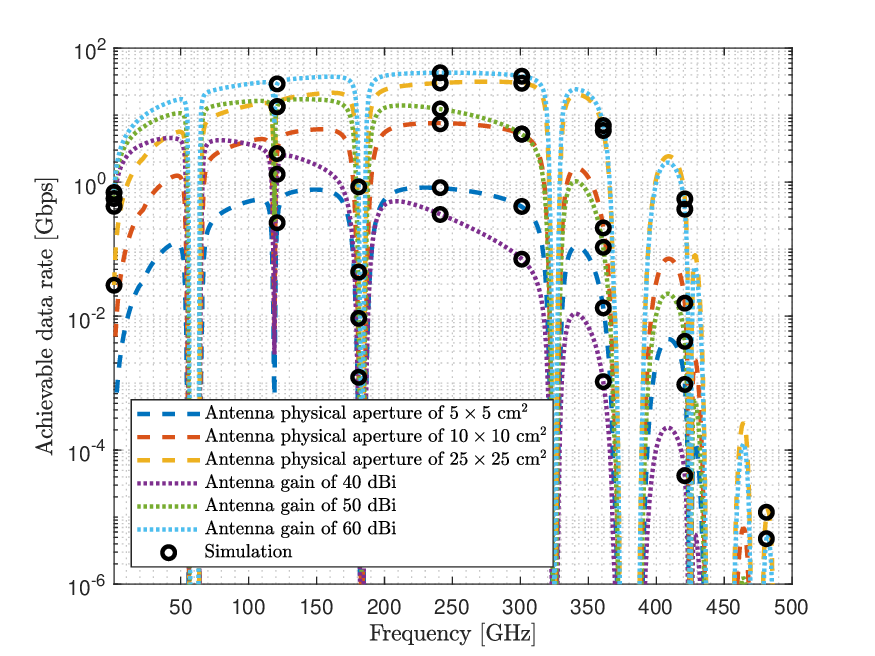}}
    \vspace{-20pt}
\caption{Achievable data rate in aerial-to-ground link for different physical apertures ($5 \times 5$ cm${}^{2}$, $10 \times 10$ cm${}^{2}$, and $25 \times 25$ cm${}^{2}$) and gains (40 dBi, 50 dBi, and 60 dBi) under humid atmospheric conditions.}
\label{fig.5.0.005}
\end{figure}

The achievable date rates in an aerial-to-ground link under humid atmospheric conditions for different physical aperture sizes ($5 \times 5$ cm${}^{2}$, $10 \times 10$ cm${}^{2}$, and $25 \times 25$ cm${}^{2}$) and array gains (40 dBi, 50 dBi, and 60 dBi) are shown in Fig. \ref{fig.5.0.005}. {Here, the fading parameters are taken as $\alpha = 6$ and $\mu = 5$, whereas standard deviation of the orientation fluctuation is assumed as $\sigma_{\theta} = 0.001^\circ$.} 
Similar to Fig. \ref{fig.5.0.004}, it is observed from the figure that the simulation curves are tightly upper bounded by the theoretical curves. 
Also, it can be inferred from the figure that both enlarging the physical aperture and increasing array gain improve the system performance. 
In addition, satisfactory data rates are not achievable above 450 GHz operating frequency, as also observed in Fig. \ref{fig.5.0.004}, even though the aerial-to-ground communication link has much shorter propagation distance. This stems from the severe absorption loss due to the dense atmosphere at low altitudes. 

\begin{figure}[!t]
\centering
 	\resizebox*{1\linewidth}{!}{\includegraphics{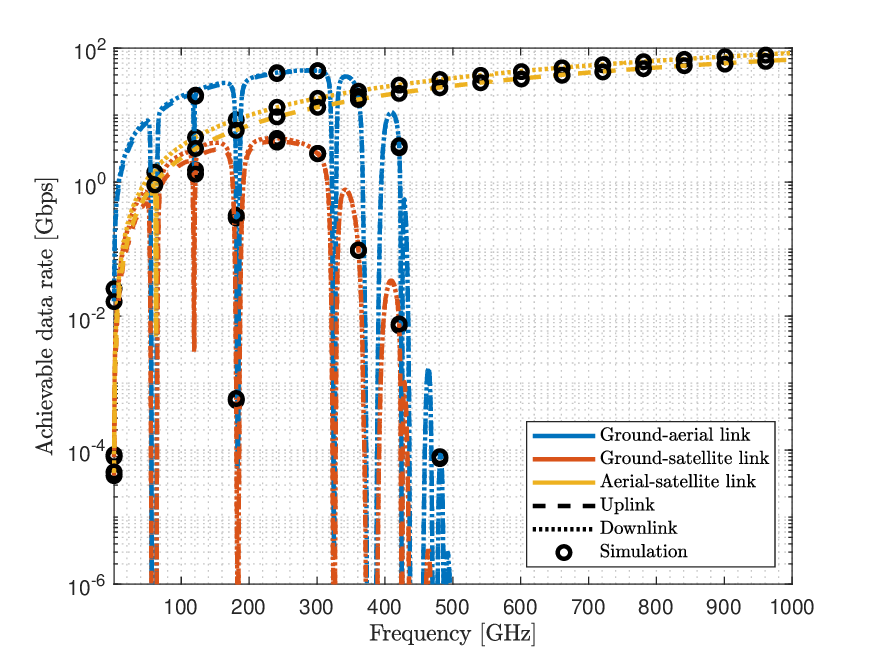}}
    \vspace{-20pt}
\caption{Achievable data rate in uplink/downlink ground-aerial, ground-satellite, aerial-satellite links for fixed physical aperture of $25 \times 25$ cm${}^{2}$.}
\label{fig.5.0.006}
\end{figure}

The uplink/downlink achievable data rates in ground-aerial, ground-satellite, and aerial-satellite links under humid atmospheric conditions are presented in Fig. \ref{fig.5.0.006}. {Here, arrays with fixed physical aperture of $25 \times 25$ cm${}^{2}$ are considered, and fading and pointing error parameters are taken as $\alpha = 6$, $\mu = 5$, and $\sigma_{\theta} = 0.001^\circ$.} 
It is deduced from the figure that ground-aerial and ground-satellite links experience frequency-selective nature of the atmosphere as the propagating electromagnetic waves pass through the lowest layers of the atmosphere in which the absorption occurs frequently. The variation between the curves for ground-aerial and ground-satellite links originates from the longer propagation path and increased free-space loss. On the contrary, aerial-satellite link has negligible frequency selectivity since it is almost not affected by the absorption owing to the very thin atmosphere {above the troposphere}. It can be concluded that data rates higher than 10 Gbps are achievable in both ground-aerial and aerial-satellite links since either the free-space loss or the absorption determines the performance in these links, where ground-satellite link suffers from both the free-space loss and absorption loss. This suggests that the utilization of mmWave/THz frequencies in NTNs can be a strong enabler for future communication systems as it reduces the attenuation owing to the multi-layer structure. \footnote{{It is important to emphasize here that low earth orbit satellites (at an altitude of 700 km) are considered in Fig. \ref{fig.5.0.006}. For satellites at higher altitudes (e.g. medium earth orbit or geostationary orbit satellites), the links distance is significantly longer than that in the low earth orbit case. Hence, it can be readily said that the received power and the achievable data rate are severely reduced due to increased free-space loss (longer link distance).}} Additionally, slight differences are observed between the uplink and downlink curves, which stems from the disparity between the upwelling and downwelling brightness temperatures (see Fig. \ref{fig.5.0.003}). 

\begin{table}[!t]
\centering
{
\caption{Array Configuration Parameters for Fig. \ref{fig_R1_3_1}}
\label{table_array_comp}
\resizebox*{1\linewidth}{!}{
\begin{tabular}{|l|c|c|}
\hline
Antenna type (HPBW \& max. gain) & Array size \& spacing & Array gain \\ \hline
Lens ($5.5^\circ$ \& 29.63 dBi) \cite{10287222} & $11 \times 11$ \& $10 \lambda$ & 50.45 dBi \\ \hline
Horn ($22^\circ$ \& 17 dBi) \cite{8608702} & $47 \times 47$ \& $5\lambda$ & 50.44 dBi \\ \hline
Cuboid ($30.2^\circ$ \& 14.5 dBi) \cite{9998671} & $63 \times 63$ \& $1.5 \lambda$ & 50.48 dBi \\ \hline
\end{tabular}}}
\end{table}

{
\begin{figure}[!t]
\centering
 	\resizebox*{1\linewidth}{!}{\includegraphics{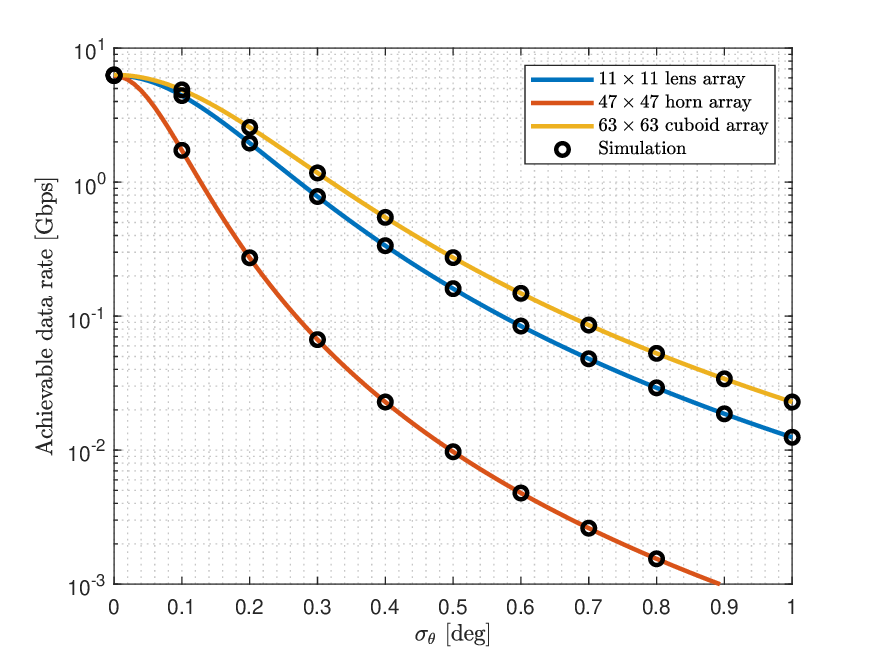}}
    \vspace{-20pt}
\caption{{Impact of jitters on the achievable data rate for different antenna array configurations.}}
\label{fig_R1_3_1}
\end{figure}

In Fig. \ref{fig_R1_3_1}, the achievable data rate is illustrated for different antenna array configurations, namely, lens antennas, horn antennas, and cuboid antennas. Here, aerial-to-ground communication link is considered at 300 GHz operating frequency for clear sky (negligible weather condition impact) and humid atmospheric conditions. The HPBW, maximum element gain, physical size of the antenna element, and the array size are given in Table \ref{table_array_comp} together with the obtained array gain of approximately $50.4$ dBi. As can be deduced from the achievable data rates, cuboid and lens arrays are more robust against pointing errors compared to horn arrays. The reason is that horn array has a much narrower main lobe characteristics compared to others, making it more susceptible to misalignment. As a result, it can be concluded that not only the array gain but also the type of antenna through which this gain is achieved is important.
}

{



}

\begin{figure}[!t]
\centering
 	\resizebox*{1\linewidth}{!}{\includegraphics{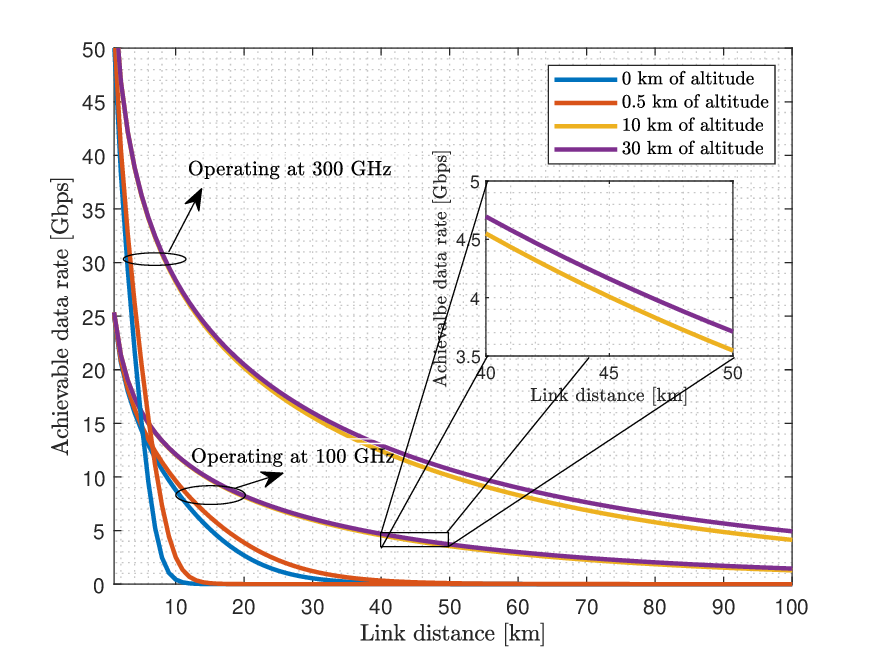}}
    \vspace{-20pt}
\caption{Achievable data rate in intra-layer communication on the ground and aerial altitudes for fixed physical aperture of $10 \times 10$ cm${}^{2}$ under humid atmosphere.}
\label{fig.5.0.007}
\end{figure}

In Fig. \ref{fig.5.0.007}, the achievable data rates in intra-layer communication under humid atmospheric conditions are shown with respect to the link distance at operating frequencies 100 GHz and 300 GHz. Here, the transmitter and receiver with fixed physical apertures of $10 \times 10$ cm${}^2$ are located at altitudes of 0 km, 0.5 km, 10 km, and 30 km, {and the link margin is assumed as $L_{\rm mar} = 6$ dB to take fading and pointing errors into account}. For link distances up to a few km at lower altitudes, enhanced data rates are achievable at 300 GHz operating frequency owing to wider bandwidth (2\% of the operating frequency) and {high array gain due to fixed physical aperture}. However, for larger link distances at lower altitudes, the absorption and free-space loss grow significantly at 300 GHz operating frequency, and the system performance deteriorates. As a result, 100 GHz operating frequency outperforms 300 GHz for longer ranges at low altitudes. {On the other hand, at 10 km and 30 km of altitude, 300 GHz operating frequency offers enhanced data rates. This arises from the absorption loss being insignificant at higher altitudes (see Fig. \ref{fig.5.0.001}), hence, the free-space loss becomes the only limiting factor, which is compensated for with the fixed physical aperture.}
Therefore, it can be said that utilizing higher frequencies enhances the achievable data rate in communication links at higher altitudes. 

\begin{figure}[!t]
\centering
 	\resizebox*{1\linewidth}{!}{\includegraphics{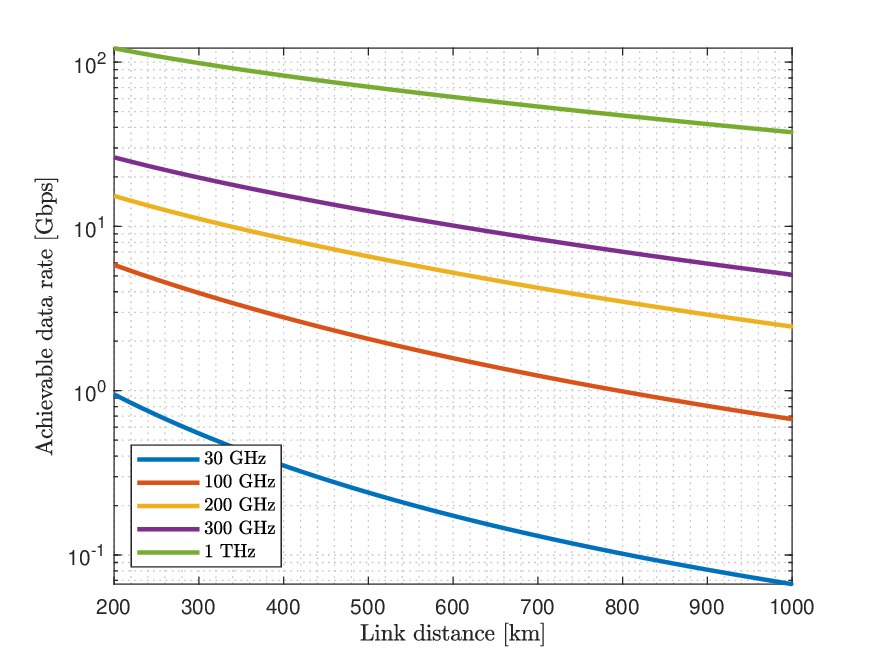}}
    \vspace{-20pt}
\caption{Achievable data rate in inter-satellite link between two low earth orbit satellites equipped with arrays of $25 \times 25$ cm${}^{2}$ physical aperture operating at 30 GHz, 100 GHz, 200 GHz, 300 GHz, and 1 THz operating frequencies.}
\label{fig.5.0.007.5}
\end{figure}

The achievable data rate in an inter-satellite link between two low earth orbit satellites orbiting at 700 km of altitude and equipped with arrays of $25 \times 25$ cm${}^{2}$ physical aperture is shown in Fig. \ref{fig.5.0.007.5} for 30 GHz, 100 GHz, 200 GHz, 300 GHz, and 1 THz operating frequencies {with $L_{\rm mar} = 6$ dB}. It can be observed from the figure that for all operating frequencies, the achievable data rate reduces for longer transmission distance because of the increasing free-space loss. In addition, the achievable data rate significantly enhances for higher operating frequency. This can be attributed to the increasing array gain and the wider available bandwidth at higher frequencies. Since the communication is established above 100 km of altitude in inter-satellite links, the atmospheric absorption is negligible (see Fig. \ref{fig.5.0.001}). {Here,} the dominant loss factor is the free-space loss which can be compensated for {by using} fixed physical aperture arrays. This allows improved performances in inter-satellite links owing to the utilization of high frequencies.

\begin{figure}[!t]
\centering
 	\resizebox*{1\linewidth}{!}{\includegraphics{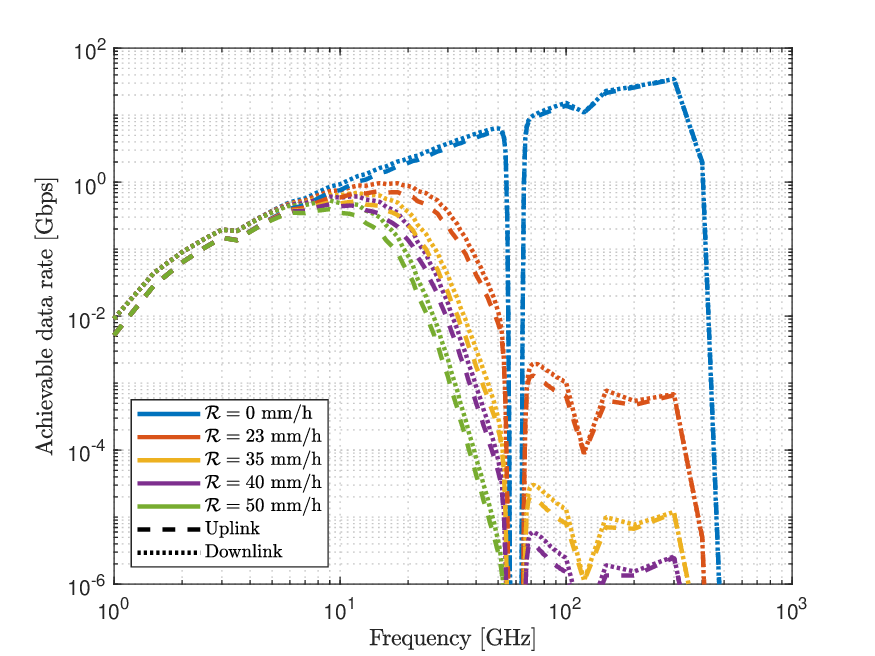}}
    \vspace{-20pt}
\caption{Impact of rainfall rate on the achievable data rate in uplink/downlink ground-aerial communication with $d_{\rm eff} = 5$ km and circularly polarized array of $25 \times 25$ cm${}^{2}$ physical aperture under humid atmosphere.}
\label{fig.5.0.008}
\end{figure}

The achievable data rate in uplink/downlink ground-aerial communications under humid atmosphere are provided in the presence of rain for $\mathcal{R} = 0, 23, 35, 40, 50$ mm/h and $d_{\rm eff} = 5$ km in Fig. \ref{fig.5.0.008}. In this figure, it is considered that circularly polarized arrays with fixed physical apertures of $25 \times 25$ cm${}^{2}$ are used {and that the link margin is $L_{\rm mar} = 6$ dB}. As can be inferred, the rainy weather does not have significant impact on the system performance below $\sim 6$ GHz.\footnote{{The slight ripple-like variations observed below 6 GHz stem from the use of the floor operator in calculating the number of antennas within a fixed physical aperture size of the array, and they become noticeable due to the logarithmic scale of the x-axis.}} As the frequency increases, higher rainfall rate leads to a poorer performance. At some frequencies, the absorption peaks {(e.g. at 60 GHz)}, form transmission windows.\footnote{Please note here that the influence of some of the previously observed absorption peaks in the humid atmosphere are not present in Fig. \ref{fig.5.0.008}. This is due to the lack of samples at high frequencies in \cite[Table V]{ITU_rain}.} Furthermore, satisfactory performance cannot be achieved at operating frequencies higher than $\sim 400$ GHz under rainy weather conditions.

\begin{figure}[!t]
\centering
 	\resizebox*{1\linewidth}{!}{\includegraphics{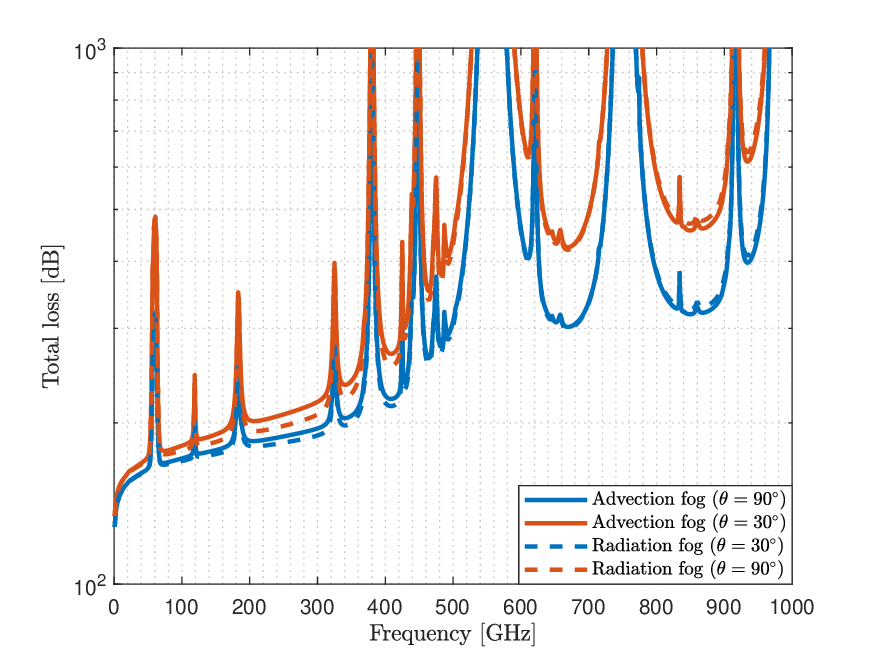}}
    \vspace{-20pt}
\caption{The total loss in ground-to-aerial link for visibility range $V = 50$ m, different elevation angles $\theta = 90^\circ, 30^\circ$ and fog types.}
\label{fig.5.0.009}
\end{figure}

{In Fig. \ref{fig.5.0.009}, the impact of cloud/fog on the total loss in ground-to-aerial link is presented for a visibility range of $V = 50$ m {with $L_{\rm mar} = 6$ dB} and different elevation angles $\theta = 90^\circ,30^\circ$ and fog types. Here, {the total loss includes all large- and small-scale effects in Section II, and} \cite{ITU_cloudfog} is followed for the calculation of the cloud/fog-induced attenuation. It can be deduced from the figure that cloud/fog formations cause significant attenuation which results in severe total loss values. Also, in this figure, several transmission windows are observed owing to the distinct absorption peaks at {60 GHz, 120 GHz, 180 GHz, 330 GHz, etc}. It can also be inferred from the figure that the advection fog results in higher total loss at lower sub-THz frequencies, whereas the radiation fog leads to slightly higher total loss values at upper sub-THz frequencies. }

\begin{figure}[!t]
\centering
 	\resizebox*{1\linewidth}{!}{\includegraphics{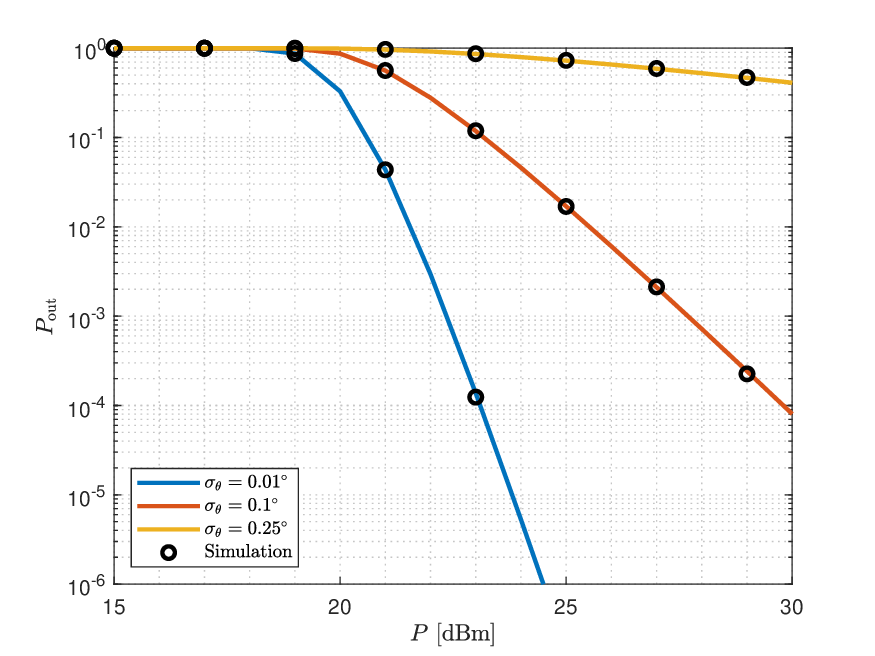}}
    \vspace{-20pt}
\caption{{Outage probability at 300 GHz in ground-to-aerial communication system.}}
\label{fig.5.0.009_outage}
\end{figure}

{
In Fig. \ref{fig.5.0.009_outage}, outage probability curves are illustrated for a ground-to-aerial communication system at 300 GHz operating frequency in clear sky and humid atmospheric conditions. In this setup, the system bandwidth is not divided into sub-bands because of the approximately flat behaviour of both the absorption loss (see Fig. \ref{fig.5.0.001}) and the noise characteristics (see Fig. \ref{fig.5.0.003}) in close proximity to the 300 GHz operating frequency. Here, $63 \times 63$ cuboid antenna arrays are used at both transmitter and receiver, and the fading parameters are assumed as $\alpha = 5$ and $\mu = 6$. The outage probability curves are shown for $\gamma_{\rm th} = 0$ dB and jitter standard deviation of $\sigma_{\theta} = 0.01^\circ, 0.1^\circ, 0.25^\circ$, where it can be inferred that higher jitter standard deviation results in a significant performance loss. In a mmWave/THz NTN setup, it is more likely to observe a favorable channel characteristics in the fading distribution sense, i.e. high values of $\alpha$ and $\mu$, due to vertical and highly directional communication. In this case, the pointing error is of critical importance as its characteristics dominate the performance.}

{
\subsection{Design Guidelines}

As thoroughly examined in the previous subsections, numerous parameters have significant impact on the received power and noise characteristics, thus on the achievable data rate. It is crucial to account for these relationships and tradeoffs while designing a mmWave/THz NTN link. The most important points in the design of such a system are listed as the guidelines as follows:
\begin{itemize}
    \item Depending on the link setup, the most destructive factor in the achievable data rate expression in \eqref{Eq.5.0.0.012_TVT2} is either the free-space loss or absorption loss. {To overcome the excessive loss levels, antenna arrays with physical apertures of $10 \times 10$ cm$^{2}$ or $25 \times 25$ cm$^{2}$ can be used.} In the design of a communication link within the NTN architecture, these dominant factors must be considered when determining favorable frequency bands, noise characteristics, and the required antenna array size. 
    \item In communication links that include the lower layers of the atmosphere (low-altitude aerial-aerial, ground-aerial, or ground-satellite links), the altitude-dependent absorption loss and weather conditions (e.g., rain and cloud/fog) have the most critical impact. Hence, the frequency bands labeled I-IV (which align with transmission windows around 60 GHz, 120 GHz, 180 GHz, and 330 GHz) are recommended. As demonstrated, data rates in other bands, specifically above 435 GHz in humid ground-to-satellite links and above 450 GHz in aerial-to-ground links, are severely degraded regardless of array gain (see Fig. \ref{fig.5.0.004}).
    \item For high-altitude aerial-aerial, aerial-satellite, or inter-satellite communication links, the atmosphere along the propagation path is thinner, resulting in negligible absorption loss. Here, the most significant destructive parameter is the free-space loss dictated by the link distance $d$. Since the loss factors in these altitudes are not frequency-selective, utilizing higher frequencies (up to 1 THz) is highly advantageous under fixed physical aperture assumption. For instance, in LEO inter-satellite links, switching to 1 THz can support robust multi-gigabit data rates for coverage distances up to 1000 km, significantly outperforming lower mmWave bands (see Fig. \ref{fig.5.0.006}, \ref{fig.5.0.007}, and \ref{fig.5.0.007.5}). For orbits of much higher altitudes than LEO (e.g., MEO or GEO), the free-space loss scales massively and becomes the ultimate limiting constraint.   
    \item While high-gain arrays enhance the achievable data rate, their highly directive pencil-beams become exceptionally susceptible to misalignment. To maintain reliability, the system design must account for pointing errors. Furthermore, careful selection of the antenna element type is required {since the statistics of the pointing errors depend on the antenna and array design. Both the selected antenna's gain and the number of antennas in the array to achieve the desired array gain determine the width of the main lobe of the radiation pattern. This inevitably influences the susceptibility of the system to the orientation fluctuations (jitters).} For example, cuboid and lens arrays are substantially more robust against orientation fluctuations than horn arrays (see Fig. \ref{fig_R1_3_1}) {as the width of the main lobe is much narrower in the horn array configuration considering the physical dimensions of the horn array, gain of a single element, and the number of antennas in the array}.
\end{itemize}

}

\section{Conclusion}

{In this work, the feasibility of utilizing THz frequencies in the NTN architecture is examined {by considering the link budget analysis}. First, the models for the main channel effects, including the free-space loss, absorption loss, weather-dependent losses, polarization mismatch and feeder losses, fading and pointing errors, are presented. Afterwards, by taking the varying noise characteristics in uplink and downlink communications due the brightness temperature and absorption into consideration, the noise power and the received signal power are found. Based on the aforementioned channel effects and noise characteristics, the achievable data rate is obtained for various communication scenarios within the NTN architecture. The results have shown that high loss levels can be compensated for by using fixed physical aperture arrays under dry atmospheric conditions. In addition, it is illustrated that the absorption and weather events have significant impacts only at low altitudes, making the THz band advantageous especially for the links at higher altitudes. Leveraging the multi-layer structure of the NTN architecture, both the free-space loss values in aerial-satellite links due to long distance and the absorption loss values in ground-aerial links due to denser atmosphere can be tolerated with high-gain directional antennas/arrays, enabling multi-gigabit links while connecting the ground, air, and space as targeted in the next-generation communication systems.}

\bibliographystyle{IEEEtran}
\bibliography{main}


\end{document}